# Population polarization dynamics and next-generation social media algorithms

N.F. Johnson[1], P. Manrique[1], M. Zheng[1], Z. Cao[1], J. Botero[1], S. Huang[1], N. Aden[1], C. Song[1], J. Leady[2], N. Velasquez[3], E.M. Restrepo[3,4]

[1]*Complex Systems Initiative & Physics Department, University of Miami, FL 33126, USA*
[2]*Mendoza College of Business, University of Notre Dame, IN 46556, USA*
[3]*Department of International Studies, University of Miami, Coral Gables, FL 33126, USA*
[4]*Department of Geography, University of Miami, Coral Gables, FL 33126, USA*

**We present a many-body theory that explains and reproduces recent observations of population polarization dynamics, is supported by controlled human experiments, and addresses the controversy surrounding the Internet's impact. It predicts that whether and how a population becomes polarized is dictated by the nature of the underlying competition, rather than the validity of the information that individuals receive or their online 'bubbles'. Building on this framework, we show that next-generation social media algorithms aimed at pulling people together, will instead likely lead to an explosive percolation process that generates new pockets of extremes.**

A fascinating debate has opened up about how the Internet impacts societal polarization *[1-8]* and the role of potentially fake news *[6]*. One might argue, as many commentators do, that social media provide an array of fragmented information sources, and that such fragmentation enhances political polarization. However, Boxell et al.'s recent empirical study casts doubt on the belief that the Internet and social media drive polarization *[1]*, with people who tend to use the Internet and social media the least showing high increases in polarization. King et al. *[7,8]* recently elucidated the power of media information in terms of activating people to express themselves. Population polarization is of great practical importance since it can manifest itself in the stories that individuals choose to share *[3,4]*, their political affiliation and how they vote *[5,9]*, and also how society views scientific findings, e.g. Darwinian evolution versus intelligent design, and climate change *[10]*. Even on issues for which there is no conceivable counter-evidence, there can be a surprisingly large number of people who take the 'anti-crowd' viewpoint, e.g. the many people who believe the world is flat and attended the 2017 Flat Earth International Conference *[11]*. Within the sub-population of professional scientists, there is also a non-zero 'anti-crowd' that are skeptical about global warming *[12]*.

The challenge of quantifying when and how population polarization develops over time (Fig. 1B,D,E,G,I,K) and the role of external information, is actually a central one for many disciplines -- from physical and chemical systems *[13]* to social, economic and political domains *[14-18]*. Foundational works in economics by Arthur *[19]*, in physics by Halpin-Healy *[20,21]*, in political science by Lazar *[4, 16, 17]*, and in psychology by Forsyth and Lewin *[22, 23]* and others *[24-26]*, suggest that addressing this challenge requires developing minimal, generative theories of a population's many-body, out-of-equilibrium dynamics – even though any such theory may be a cartoon version of the real-world socio-economic and political system *[4,16,19,20,21,24-31]*. Here we present such a theory (Fig. 2) which quantifies conditions for population polarization and its dynamics; is backed up by controlled human laboratory experiments (Fig. 2B and Methods and Data); and which reveals the observed polarization dynamics as being driven by a many-body collective process that builds crowds and anti-crowds at the extremes (Fig. 1A,C,F,H,J, Fig. 2B). Each individual can adapt his/her behavior (i.e. continuous $p$ value where $0 \leq p \leq 1$) in time according to its success, and hence the heterogeneity within the population (i.e. distribution $P(p,t)$) evolves over time. Most existing models maintain the heterogeneity distribution as fixed while allowing the location or connectivity to change in some spatial network, or they consider individuals having a discrete, binary state (e.g. 0 or 1, spin up or spin down) which can fluctuate while the network location or connections remain fixed *[25,29-31]*. Such models cannot reproduce or explain the time-dependent evolution and shapes of *P(p,t)* observed in Figs. 1B,D,E,G,I,K and 2B. We also predict that the addition of next-generation social media enhancements (Fig. 2C,D) akin to those recently proposed by Facebook, will generate a new form of explosive percolation *[32]* within the population (Fig. 3). Although these algorithms can eventually connect together the majority of the population, the process will likely generate new pockets of isolated extremes.

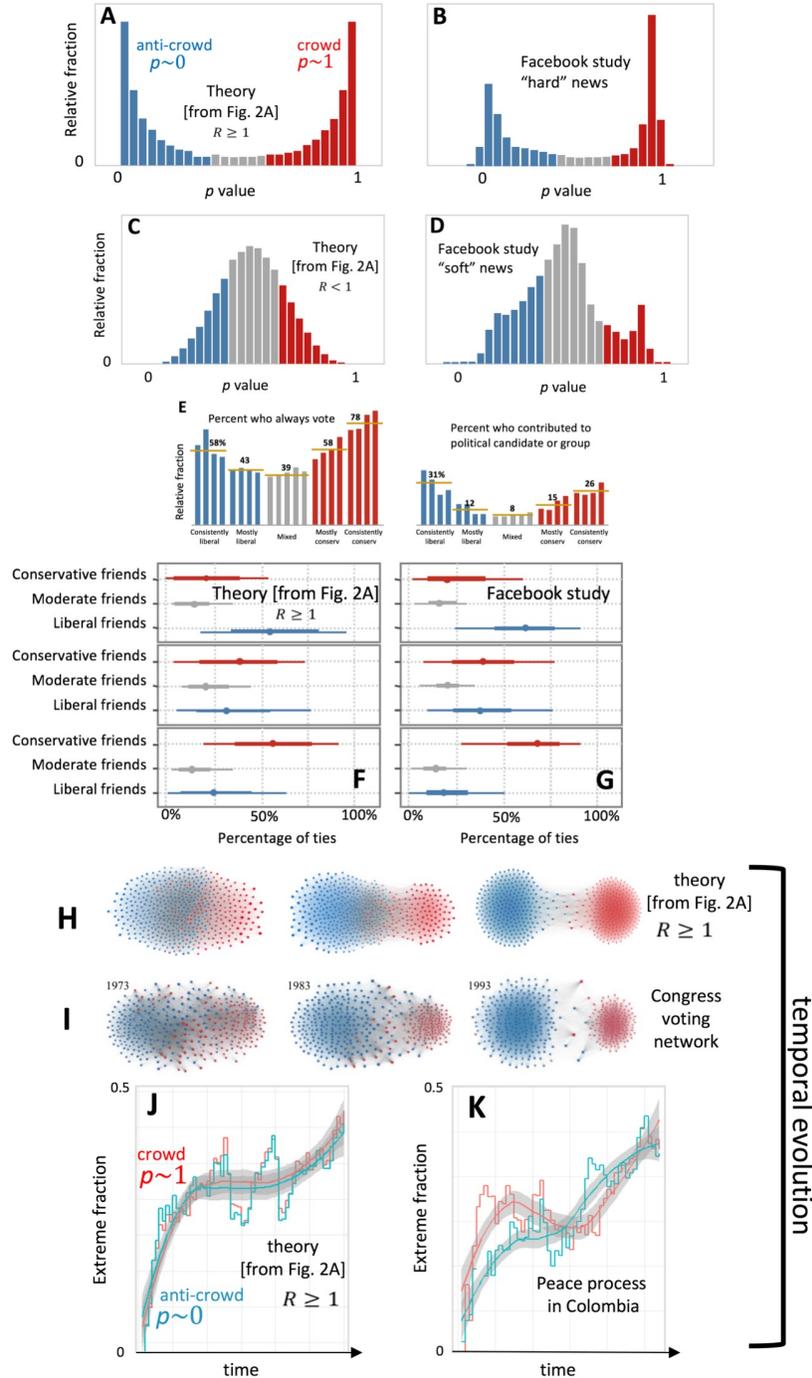

**Fig. 1: Real-world polarization and theory. A-D:** Heterogeneity distribution *P(p,t)* showing the relative fraction of population having a particular *p* value (i.e. re-scaled alignment using the terminology of Ref. 3. A,C show our theory for $R \geq 1$ and $R < 1$ respectively, with $R$ near 1. B,D: empirical data from Ref. 3 for "hard" and "soft" news respectively. E: other empirical examples from Ref. 5. F,G: theory and empirical data *[3]* respectively, for the percentage of ties per individual type. H,I: theory and empirical data respectively for the time-dependent voting network of Ref. 9. J,K: theory and our empirical data (see Methods and Data) respectively for the relative fraction of the two extremes in the Colombian peace process as they evolve over time during 2016-2017.

Figure 1 shows the favorable comparison between our minimal many-body theory (from Fig. 2A) and real-world data, not only for snapshots and time-averages where polarization arises (Fig. 1B,E,G) and where it does not (Fig. 1D), but also for the dynamical evolution prior to polarization (Fig. 1I,K). Figure 2B shows similarly good agreement with the empirical dynamical evolution, but now for our controlled laboratory experiment on a smaller population (see Methods and Data). Our theory (Fig. 2A) builds on the formal structure introduced by economist W.B. Arthur *[19]* to explore how humans operate when there is no obvious deductive, rational solution for how they each should act *[19]*. We consider an arbitrary number $N$ of heterogeneous, adaptive individuals who are going through life in pursuit of a piece of a common 'pie' of size $L$, i.e. a resource or benefit which is neither so small that nobody can access it, nor so large that there is no need to try. We do not need to complicate the discussion by specifying what that resource or benefit is: it could be material in a socioeconomic context, or reputation, prestige or influence in some sociopolitical context, and the population could be a particular sector of society or an entity, or a geographic population such as a city, state or country etc.

Each individual in the population sporadically receives an information update $I$ via news or social media source, concerning economic, political and/or social prospects for the population. Despite the age of fragmented media *[8]*, the landmark study of King et al. *[7]* shows that such common information ($I$) remains more influential, relevant and connected to a broad cross-section of the population than might otherwise have been thought. The collective dynamics turn out to be insensitive to whether $I$ is *presented* as good (which we encode as 1, suggesting that individuals should step up their efforts to chase $L$) or bad (which we encode as 0, suggesting that individuals should conserve their efforts until the next $I$ appears) or if $I$ is true or fake, or anything in between. In the language of physics, $I$ acts as a time-dependent field. We also do not need to assume this is some zero-sum game or that $L$ or $N$ don't change in time, but for simplicity we focus on $L/N \sim 0.5$ since this is the "glass half-full-half-empty" scenario meaning that there is no a priori bias in terms of how individuals should act, and hence individuals need to strategize as to when to step up their efforts (e.g. spend extra energy, money) in pursuit of $L$. Each individual uses his/her current strategy $p$ to decide how to interpret $I$ (i.e. good or bad, true or fake) and hence decides whether it is now worth stepping up to chase $L$ and hence risk wasted efforts or not. Each strategy $0 \leq p \leq 1$ is the probability that the individual takes $I$ at face value and hence acts accordingly: if $I=1$, then with probability $p$ the individual subsequently chases $L$ on the basis that prospects are good and so it is worth the risk. If $I=0$, then with probability $p$ the individual does not risk wasted efforts chasing $L$. During the time until the next $I$ arrives and the process repeats, if the total number of individuals actively chasing $L$ is less than $L$, those individuals consider themselves as having won and hence assign a reward $R \sim 1$ to the reliability score of their particular $p$. If it is more than $L$, those individuals consider themselves as having lost and hence assign a unit penalty to their $p$'s reliability score – or we can equivalently consider individuals as gradually losing patience with their $p$ and hence the reliability score drifts steadily downwards. If an individual's $p$ has a reliability score that falls below some negative value $d$, he/she adapts by changing their $p$ to a new value either side of the old one (without bias). This setup builds directly on prior work in the economics and physics communities *[33-40]*.

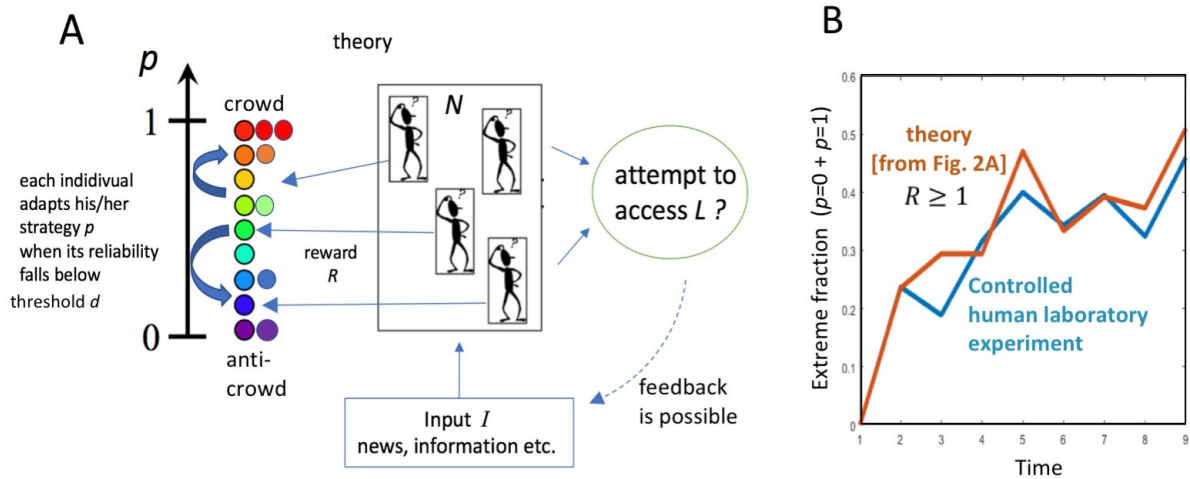

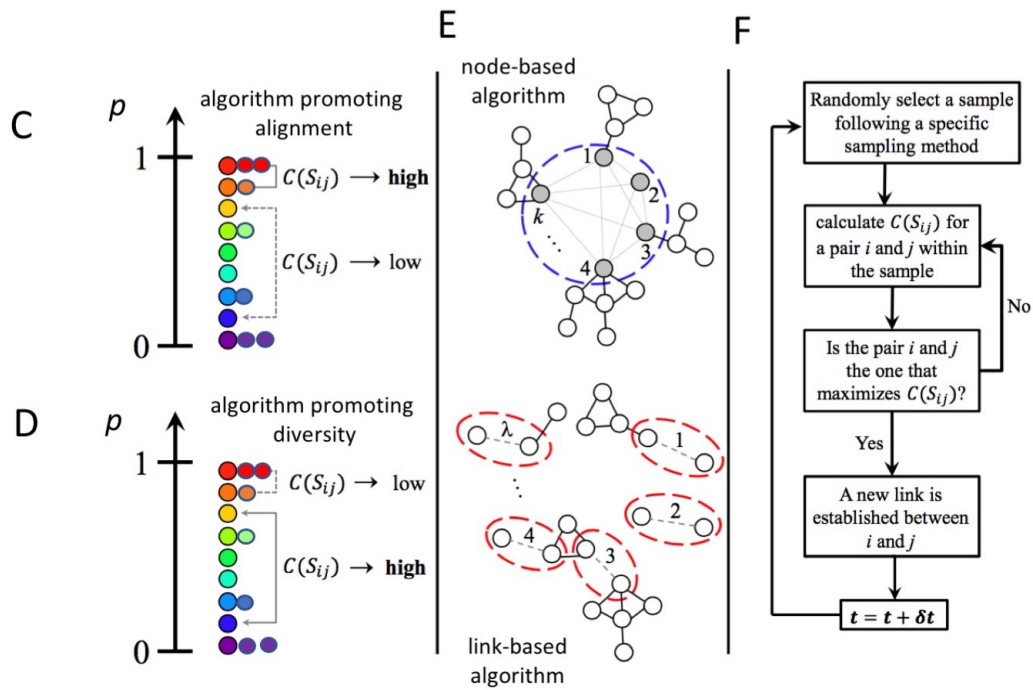

**Fig. 2: A:** Theory describes an evolving population of boundedly-rational individuals *[19]* pursuing individual gain in pursuit of some common resource or benefit of size *L*. Each sporadically receives information *I* and decides how to act using their individual *p* value, and changes it when it becomes unreliable. The distribution of *p* values at time *t* is the heterogeneity distribution *P(p,t)*. **B:** Evolution of polarization in our controlled laboratory experiment (blue line, see Methods and Data) and theory (orange line), with $R \geq 1$. **C,D:** Two next-generation social media algorithms which actively connect individuals with the aim of promoting alignment and diversity respectively. See Methods and Data for meaning of $C(S_{ij})$. **E:** Our results are insensitive to how these two algorithms are implemented, i.e. choosing $k$ individuals (nodes) or $\lambda$ links. **F:** Example flow-chart of these two algorithms.

Numerical simulation of our theory (Fig. 2A) shows that the heterogeneity distribution *P(p;t)* evolves dynamically in a highly non-trivial way. To the extent that an individual's *p* value captures his/her ideology, *P(p;t)* provides a prediction for how the population will collectively act at any given time *t* in terms of, for example, sharing of particular material online or voting. This is supported by the good agreement shown in Figs. 1 and 2B. Our theory reproduces the details of the observed ties from Ref. 3 (Fig. 1F,G) simply by tracking through time the people with whom each individual shares a similar *p* value. Most importantly, our theory captures the dynamics of the polarization extremes exhibited by the U.S. congressional voting data (Fig. 1H,I), the Colombian peace process data (Fig. 1J,K) and our controlled laboratory experiments (Fig. 2B). In particular, it reproduces the novel plateau-like structure that emerges in Fig. 1K and is hinted at in our experiments in Fig. 2B. This dynamical bottleneck represents a temporary slowing-down of the polarization process, which in turn means that *there is an extended period of time during which the polarization changes little, and hence during which interventions could be made to try to reverse the polarization process*. This is the first report of such a societal polarization bottleneck that we are aware of. Our theory also captures the U and inverted U shapes of the steady-state and snapshot results for *P(p;t)*. Whether the population develops a steady-state polarized U-shape as in Fig. 1A, is dictated in our theory by the reward to penalty ratio *R*. A large enough *R* (i.e. $R \geq 1$) prevents adaptation from being too fast and drives the formation of similarly sized groups around the extremes *p*~1 and *p*~0 (crowd and anti-crowd). The closer individuals are to *p*~1 and *p*~0 respectively, the more likely they are to take opposite positions from each other. When *R*<1 *[35,38,39,37,34,36,40]* an inverted U emerges (Figs. 1C,D). This provides a parsimonious explanation for the findings of Ref. 3 since "hard" news (Fig. 1B) is directly relevant to the current socioeconomic climate and hence will be considered as relevant to an individual in deciding whether to pursue *L* at that moment, and hence *R* would be larger which is consistent with $R \geq 1$ (Fig. 1A). "Soft" news (Fig. 1D) is not directly relevant to the current climate and hence will play less of a role in whether an individual decides to pursue *L* at that moment: hence *R* would be smaller, consistent with *R*<1 (Fig. 1C). In stark contrast to the key role of *R*, the emergence of the polarized U-shape is *not* sensitive to the other parameters in the model, i.e. individuals' threshold value *d*, the size of the change of *p* upon adaptation, the absolute values of *L* and *N* at fixed *L/N*=0.5, or the nature of the information *I*. It is also *not* sensitive to (1) whether the information *I* is fabricated to provide artificially high levels of 'good' news (e.g. all 1's) or 'bad' (e.g. all 0's), or whether it is true or false, or whether it is released in a certain sequence; (2) whether we include population heterogeneity from the outset by randomly assigning *p* values, or set them all equal to 0.5 to mimic maximum initial uncertainty of all individuals; (3) the origin or precise nature of *I*; (4) the amount of clock-time that passes between the new arrival of information *I*; (5) the absolute value of the reward and penalty (only the ratio *R* is relevant); (6) the way in which individuals adapt and hence choose a new *p* value, as long as they do so independently; (7) whether different sectors of the society receive different *I* and hence exist in different news bubbles. As long as each bubble remains self-contained and has the same *R*, each U or inverted U will just add together and hence preserve the same shape; (8) other 'bubble' mechanisms, such as allowing individuals with similar *p* values to mimic each other. This applies even if the population gets broken up into only three bubbles. In short, while any of these factors may impact the way in which *P(p;t)* evolves and the time it takes to reach a steady-state, they do not affect whether the final *P(p;t)* is U-shaped or not. Since a U and inverted U are symmetric, even a switch in definition of what is extreme 'left' and what is extreme 'right' would not change the resulting *P(p;t)*. We also note that generalizing the characterization of people's strategies and hence ideologies from a single *p* value to a vector of attributes of arbitrary length, and/or allowing *I* to be a vector, does not change the emergence of crowds and anti-crowds at the extremes *[40]*.

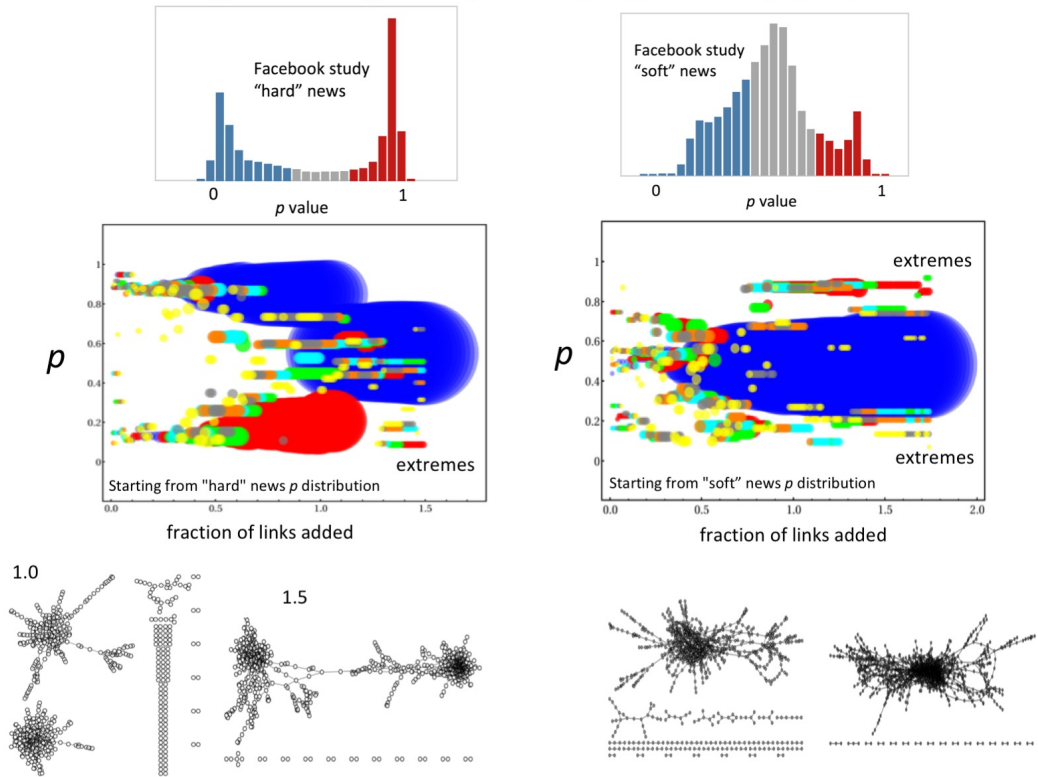

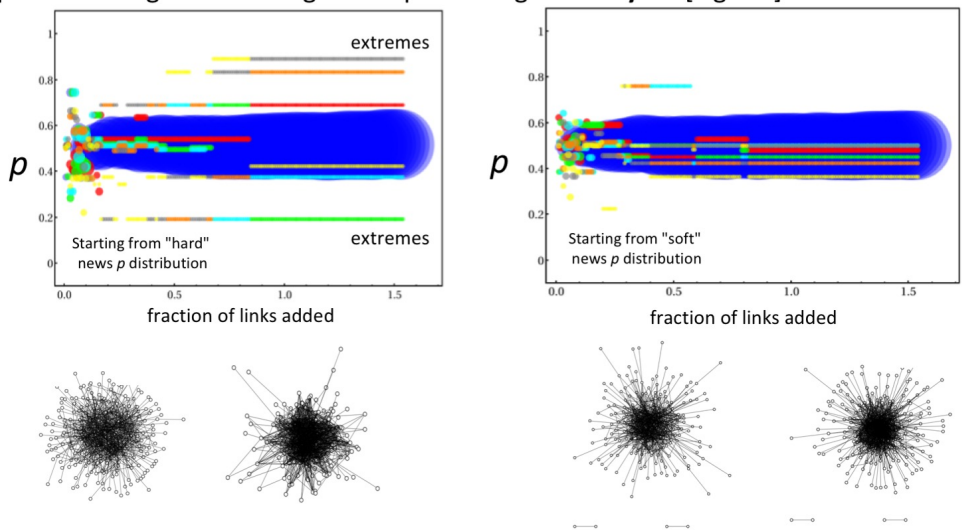

**Fig. 3: A**: Impact of adding links using next-generation algorithm promoting alignment (from Fig. 2C) on the heterogeneity distribution from Fig. 1B (left-hand side) and from Fig. 1D (right-hand side). Middle panels show circles whose center sits at the average $p$ value for each of the seven largest clusters that are formed, and whose radius is proportional to the number of individuals in that cluster. Results shown as a function of the increasing fraction of links added. Different colors are used to denote different clusters. In both cases, small isolated clusters appear with more extreme average $p$ values away from $p=0.5$ even when there is a single large cluster formed around $p=0.5$ and even though the fraction of links added is greater than 1.0 so that each individual is included in at least one link on average. Lower panels show snapshots of the actual network, at two values of the added-link fraction. The network for the "hard" news heterogeneity distribution (left-hand side) shows that even the single large cluster that exists at added-link fraction 1.5 is actually internally highly segregated. It requires only one weak link to be broken in order to fragment the cluster into two separate pieces. **B**: Same as A, but for the next-generation algorithm promoting diversity (from Fig. 2D). In this case, the main cluster forms quickly around $p=0.5$, but it leaves a number of small isolated clusters with more extreme average $p$ values that again survive until a surprisingly large fraction of links is added ($\gg$ 1.0).

Given Facebook's summer 2017 announcement of next-generation features that will actively create links between people, we now build on the above framework to explore the likely impact that such algorithms (Figs. 2C,D) will have starting from the present-day $P(p;t)$ profiles shown in Figs. 1B,D *[3]*. Since any machine-based mechanism can create links quickly once implemented, we assume a quasi-static $P(p;t) \equiv P(p)$ while these algorithms are operating. The alignment-algorithm (Fig. 2C) links together individuals based on the similarity of their *p*, and hence promotes the growth of clusters with high internal alignment. The diversity-algorithm (Fig. 2D) links together individuals based on the dissimilarity of their *p*, hence promotes the growth of clusters with high internal diversity. We have checked that our results are insensitive to exactly how these algorithms operate (Fig. 2E shows two obvious implementations in which either a candidate individual (i.e. node) or link is selected, while Fig. 2F shows the flow diagram).

Figure 3 shows that although both algorithms are designed with the intention of pulling together the population, *several unexpected and potentially undesirable features emerge as the links are added*. First, isolated residual clusters arise with an internal average *p* value well away from the population average of *p*=0.5. Indeed, for the alignment-algorithm with either the "hard" or "soft" news *P(p)*, and for the diversity-algorithm with the "hard" news *P(p)*, these isolated clusters are extreme in that their average internal *p* value is very close to the extremes *p*=0 and 1. This is particularly surprising for the alignment-algorithm in the case of "soft" news, since the "soft" news *P(p)* population starts off with an average $p\sim0.5$. Second, these extreme clusters persist even when the fraction of links is $\gg 1.0$ and hence each individual has on average at least one new link. Third, even though links are added smoothly and monotonically by the algorithms, the cluster evolution shows an abrupt temporal evolution akin to explosive percolation *[32]*. Fourth, even though at high enough link fraction $\gg 1.0$ the population is largely reunited into a single cluster, the network comprising the largest cluster for the "hard" news diversity-algorithm is still internally highly segregated (see Fig. 3A left lower panel) meaning that it requires only one weak link to be broken in order for the population to fragment into two disconnected pieces.

In addition to the mechanisms in Figs. 2C,D providing plausible representations of next-generation algorithms, there is empirical evidence that our theory in Fig. 2A provides a plausible representation of human behavior. In addition to Arthur's original real-world observation *[19]*, this evidence comprises: (1) The *controlled laboratory experiments* carried out by one of us *[33]* featuring the same setup of *N* individuals repeatedly deciding whether to access a common resource *L<N*. This controlled laboratory experiment differs from Axelrod's Prisoner's Dilemma tournaments *[41]* in that it has a formal structure, it provides a tangible incentive for the individuals, and it uses the same subjects throughout. It makes no a priori assumptions about the structure of the learning dynamics. By focusing on the choice of a *strategy* as in our theory, it also differs from the standard economics experiment in which a subject chooses a single action during a round of play *[42]*. In accordance with our theory, these experiments show that individuals are heterogeneous in terms of strategies used; that they adopt different strategies over time based on their past success; that they show a tendency over time to move away from more complex mixed or conditional behaviors toward simpler strategies given $R \geq 1$, i.e. 'always' in Ref. 33 which is akin to $p\sim1$ and $p\sim0$ and which also happen to become the most successful strategies in our theory (see later); that individuals do not manage to collectively find a more optimal solution (e.g. implicitly taking turns) but instead continually compete. A different set of experiments which also support these mechanisms, was carried out by Stanley et al. *[43]*. (2) The *controlled real-world experiments* of Konstantakopoulos et al. *[44-46]* feature *N* individuals trying to access limited common lighting *L<N* and being rewarded accordingly. Though now in a real-world office setting, the findings are again consistent with our theory. (3) The *uncontrolled real-world phenomenon* of financial trading, in particular Refs. 47,48, again confirms that, like our theory, the most popular and successful strategies turn out to be the 'always' type (e.g. momentum akin to $p\sim1$ and contrarian akin to $p\sim0$). Moreover, this was shown both for the case of humans choosing strategies when trading for themselves, and when choosing algorithms to write for machine trading. Other empirical studies with similar results include studies of drivers repeatedly deciding whether to access a potentially congested route *[49]*.

We now provide mathematics that helps explain and quantify the process of population polarization, and hence our findings in Fig. 1. We adopt three complementary perspectives which have been used successfully for physical systems: few-body, *N*-body statistical, and large-*N* dynamical. (1) *Few-body*: Since three is typically the minimum number of degrees of freedom for which complex dynamics can arise in a system, consider $N=3$ individuals (or three groups sharing a common *p* value) and three possible *p* values: 0, 0.5 and 1. Halpin-Healy *[20,21]* showed how such a 'three-body' theory in a comparable setting of conformity and dissent can provide a powerful quantitative description. Computer simulations of this $N=3$ version of Fig. 2A reproduce similar U and inverted U-shapes as for general *N*. There are $3^3=27$ possible microstates (i.e. arrangements of the $N=3$ individuals among the 3 *p* values) *[34]*, as shown in the SM. A microstate $(n_0, n_{0.5}, n_1)$ denotes $n_0$ individuals with $p=0$, $n_{0.5}$ with $p=0.5$ and $n_1$ with $p=1$, and each provides a certain average payoff per individual after accounting for rewards and penalties. For $R=1$, the highest payoff microstates are (1,1,1) which is 6-fold degenerate, i.e. 6 arrangements of the individuals 1,2 and 3, and (1,0,2) and (2,0,1) which are each 3-fold degenerate. Even though the population macrostate (U-shape as in Fig. 1A) appears static, the population continually transitions between these 12 microstates since individuals keep sporadically changing their *p* values – albeit at different rates – since there is no *p* value which is so good that it guarantees never needing adaptation. We can therefore take an average over time, to provide a theoretical prediction of the imbalance in heterogeneity. Assuming that all these microstates are visited equally, it follows that *P(p)* should have a U-shape in which *P(0)= P(1)=2.5 P(0.5)*. Since the same is true for any set of three individuals, the general *N* population will also have a U-shape. Though the ratio is larger in the continuous-*p*, large-*N* case, this finding is consistent with the simulations (Fig. 1A). Moreover as in the simulations with large *N*, individuals with $p\sim0$ and $p\sim1$ retain their *p* value for a very long time while those with $p\sim0.5$ continue to adapt most quickly. $R>1$ behaves similarly. $R<1$ favors microstates such as (0,3,0) and hence an inverted-U. (2) *N-body statistical*: Counting the probabilities for how adaptation takes individuals in and out of a specific region of *p*-space, an argument parallel to that of Ref. 36 (see SM) yields the following general expression:

$$P(p) = A \left[\frac{1}{2} - pF_N^< - (1-p)F_N^> + 2p(1-p)G_{N-1}\right]^{-1} \quad (1)$$

where explicit expressions for $F_N^<$, $F_N^>$, $G_{N-1}$ are derived in the SM, and *A* is determined by the normalization condition that *P(p)* integrates to unity. Despite their complex mathematical forms, the terms appearing in Eq. (1) have a simple interpretation that elucidates why and when population polarization emerges. Consider an individual who is merely observing this theoretical 'rat race' and hence not adding to the demand for *L*. The probability that their *p*-value wins at any given instance, is given by ½ corrected by the second and third terms on the right-hand side. The fourth term is the one that accounts for their own impact in the system, just like a driver adds to the likelihood of a jam by being present on the road or a buyer inadvertently raises the asking price of an item simply by showing interest. When it is large enough, it can cause the inverted U-shape *P(p)* to flip into a U-shape *P(p)*. (3) *Large-N dynamical*: Using an argument paralleling Ref. 37, it can be shown that *P(p;t)* in the large-*N* and continuous-time limit, satisfies a generalized diffusion equation $\frac{\partial P(p;t)}{\partial t} = \frac{\partial^2 f[P(p;t)]}{\partial p^2}$ where $f[..]$ is a functional of *P(p;t)* and hence describes the evolution of *P(p;t)* in time. Additional analysis then becomes possible using the time-dependent random-walk work of Refs. 38-40. Though *P(p;t)* cannot be obtained in closed form, the competition between the emergence of a U-shape and inverted U-shape in the long-time limit is reflected in the fluctuations (variance) given by $N \int_0^1 p(1-p)P(p;t)dp$, which has extrema when *P(p;t)* is localized around *p*=0.5 and when *P(p;t)* is localized around *p*=0 and *p*=1. The time-dependence of the polarization can be captured by a mean field-like equation for the rate-of-change of the fraction *x(t)* of the population in the crowd and anticrowd: $dx/dt = r + a(x - x_0)^2$ where $r$ is the average growth rate, and the quadratic term is a self-interaction term. We estimate $x_0 \sim \frac{1}{3}$ which is the approximate fraction of the population not in the crowd or anticrowd when *P(p;t)* is flat. For $r > 0$, there is a saddle-node ghost at $x \sim x_0$ *[50]* hence *x(t)* initially increases fast from $x(0) \sim 0$, then forms a bottleneck, plateau shape around $x(t) \sim x_0$ before accelerating again – as observed in the real population polarization during the Colombian peace process in Fig. 1K, in the controlled human laboratory experiments in Fig. 2B, and in our theory in Figs. 1J and 2B. The time spent by the population in the bottleneck can be obtained by integrating $dx/dt$ to give approximately $a\pi/\sqrt{r}$ which diverges as $r \to 0$. *This suggests that even in a population that is in the process of polarizing, there will be an extended intermediate interval of time during which intervention could be introduced to push the system polarization in the other direction.*

In the future, as new features are proposed to enhance the technology and connectivity of online media, the appropriate network connections can be added into the simulations and incorporated into the above mathematical analysis within a mean-field approximation. This will allow predictions to be made about the likely impact of new social media features on the population polarization, and hence appropriate adjustments can be made before it is launched.

**Methods and Data**
Our controlled laboratory experiment (Fig. 2B) to test the theory (Fig. 2A) was implemented by one of us with full detailed documentation available online *[33]*. The computer simulation of our theory in Fig. 2A is straightforward and builds on existing software that is freely available for Arthur's El Farol problem *[19]*. To run simulations within any browser and without requiring any programming skills, see http://ccl.northwestern.edu/netlogo/models/ElFarol. For instructions on how to modify this simulation code, see http://citeseerx.ist.psu.edu/viewdoc/download?doi=10.1.1.60.7968&rep=rep1&type=pdf.
Our data from the Colombian peace process comes from the public posts of the 218 Facebook collective profiles (i.e. groups and pages, but not private individual accounts) identified through online trawling, whose discussions were centered on the peace talks between the State and Marxist guerrillas. The data capture methodology copies exactly the approach introduced and discussed in depth in Ref. 51. The data was captured through Facebook's API for the period between September 2012 and February 2017. Each collective was coded according to its polarity (i.e. supporters, opponents, or neutral discussants of the peace policies). The logs comprise more than 4 million messages, that were processed to identify the number of unique commenters (i.e. those who posted at least once), per month, per polarity. While Schelling's segregation model (SSM) *[25,29-31]* also produces U-shapes, the underlying mechanisms and assumptions are completely different: in SSM, individuals locally try to be in the majority whereas in the present theory they are competing for *L* and do not require any local geographic connectivity; in the SSM the decision is made based on the present state of the system and is deterministic, while in the present theory the decision is stochastic based on a probability *p*; in the SSM, the individuals have no memory whereas in the present theory they record the scores of their *p* values, and also can include memory of the past without changing the conditions under which a U-shape emerges; the SSM is scale-dependent, i.e. the segregation emerges depending on how the agents choose their boundaries which implies local alignment rules, while the present theory is scale-free and independent of any geometry; the present theory can produce a U-shape or an inverted U by varying *R*, but in the SSM even if all individual agents have a strict preference for perfect integration, best-response dynamics may lead to segregation. The next-generation algorithms in Fig. 2C,D are implemented as follows: The mechanisms of link addition follow directly from the relationship among the *p* values associated to the individuals (i.e. nodes) to be linked. This is quantified by the similarity $S_{ij}$ between individual *i* and individual *j* which is defined as $S_{ij} = 1 - |p_i - p_j|$. Thus highly similar individuals are close to each other in the *p* distribution, and otherwise for highly dissimilar individuals. The mechanisms of link addition depend on the coalescence function $C(S_{ij})$. The alignment-algorithm (Fig. 2C) favors connecting similar individuals. The diversity-algorithm (Fig. 2D) favors connecting dissimilar individuals. A system following the alignment-algorithm tends to form clusters of alike individuals while diversity-algorithm tends to form clusters with unlike or complementary individuals. Hence, for the alignment-algorithm a coalescence function is defined as $C(S_{ij}) = S_{ij}$ while for the diversity-algorithm, $C(S_{ij}) = 1 - S_{ij}$. The present state, i.e. prior to any next-generation algorithms, has no additional links. At each timestep, a sample from the system is randomly selected and a new link is established between the pair of individuals that maximizes the coalescence function $C(S_{ij})$. Though the sampling can be either of individuals (i.e. nodes) or links, the evolution of the network presents similar properties.

**Acknowledgments:** NFJ gratefully acknowledges funding under National Science Foundation (NSF) grant CNS1522693 and Air Force (AFOSR) grant FA9550-16-1-0247.

**Supplementary Material (SM)**
1. Model in Fig. 2A and analytical results for *N*=3
2. Derivation of Eq. 1 and explicit expressions for the terms
3. Additional details behind the next-generation algorithm implementation